\documentclass[12pt]{iopart}
\usepackage{iopams}
\jl{6}

\begin{document}

\title{Classical Dark Matter}
\author{Mark J Hadley}
\address{Department of Physics, University of Warwick, Coventry
CV4~7AL, UK}
\ead{Mark.Hadley@warwick.ac.uk}


\begin{abstract}
Classical particle-like solutions of field equations such as general relativity, could account for dark matter. Such particles would not interact quantum mechanically and would have negligible interactions apart from gravitation. As a relic from the big bang, they would be a candidate for cold dark matter consistent with observations.
\end{abstract}

\pacs{95.35.+d, 11.90+t, 04.20.-q}
\submitted
\maketitle

\section{Introduction}
Einstein's dream of unification was to describe elementary particles as solutions of a classical field theory. Early attempts were made to find particle like solutions to General relativity and extensions of it, but spherically symmetric solutions did not correspond to the known elementary particles. General relativity itself has a family of particle-like solutions - The Kerr solution; parameterised by mass, M, and Angular momentum, $a$, which reduces to the Schwarzschild solution for $a = 0$. However, these full solutions all have singularities.

The current fashion in theoretical particle physics is to assume that quantum theory is ubiquitous and to work primarily with quantum mechanical wavefunctions, creation operators etc. Candidates for dark matter have been restricted to dark baryonic matter and more exotic particles whose wavefunctions are suggested by extensions of the standard model such as string theory. Gravitation is assumed to be a quantum phenomenon that will eventually be described by a quantum theory of gravity or by string theory. However there is no experimental evidence for quantum gravity and there is no theoretical requirement for quantum theory to be universal.

Any geometric theory of space-time, offers the possibility of both spacelike solutions that evolve with time and also solutions with non-trivial causal structure. General relativity is the obvious example of such a theory, but the results such as Geroch \cite{geroch} and Hadley \cite{hadley97} apply more generally to any geometric theory of space and time. The causal and acausal solutions are quite different in character.

The causal solutions would be hypersurfaces, possibly with non-trivial spatial topology which evolve with time. They would behave like classical objects - waves or particles. Their evolution would be deterministic. From Geroch's theorem, singularity free solutions would not be able to change topology. While this would allow scattering interactions, it would prevent reactions of the form $A+B \rightarrow C + D$ if C or D had different topology to A or B. Even reactions of the form $A+B \rightarrow C$ could only create a product that was a separable union of A and B. Particle-like solutions would have an approximately schwarzschild geometry at large distances (in the weak field limit) and would therefore experience gravitational attraction according to classical general relativity.

The acausal solutions would, by definition, allow context dependence where \emph{initial} conditions could not be fully defined without some knowledge of \emph{future }conditions. Propositions related to such space times would have the same logical structure as quantum theory - an orthomodular lattice of propositions \cite{hadley97}. The equations of quantum theory are one way to describe the evolution of such structures. As far as is known, the subspaces of a Hilbert space provide the only non-trivial representation of probabilities on an orthomodular lattice. Context dependence allows indeterminism and topology changing interactions of the form $A+B \rightarrow C +D$. Transformations under rotation would normally allow half integral values of spin because the time parameterised rotation vector field could not be extended throughout the manifold.

According to this view the particles we see are the acausal solutions because they are able to interact quantum mechanically and are not restricted to gravitational interactions. Classical particle-like solutions that interacted only gravitationally could also exist but they would be unobservable apart from their gravitational effects.

\section{Dark Matter}
Classical particles created at the big bang would contribute to cosmological models as cold dark matter. The weakness of the gravitational interactions and the absence of topology changing interactions would mean that the classical particles would have been decoupled from radiation and other forms of matter from the earliest times.


The particles would have a characteristic length scale related to their mass: $ r_m \simeq G m/ c^2$ which for a mass of 1eV is $1.3 \  10^{-63}$m, which is twenty eight orders of magnitude smaller than the Planck length. If they were compact enough to have an event horizon it would be at $2 r_m $.

The length scales involved are so small that criteria for Hawking radiation are not applicable. Hawking radiation is derived by considering a wavefunction on a fixed background spacetime. To extend the result to black holes of the order of 1eV would require the wavefunction to be defined in volumes many orders of magnitude less than the Planck length scale. Either the size and structure of the elementary particles, or their gravitation fields will be large compared with the background curvature of space due to the classical particle. Although it would be wrong to describe these classical particles as primordial black hole relics, the limitations on evaporation of black holes \cite{carr_94} are relevant.

Even small mass classical particles could have sufficient density to account for dark matter halos in galaxies. Although the exclusion principle limits the total mass of neutrinos in a galaxy \cite{tremaine_gunn}, that applies to fermions with spin-half and would not apply to classical particles. Some very specific classical geometric models can give transformation properties of a spinor under rotations, but these models require either a lack of time orientability \cite{hadley2000} (which has been suggested as an explanation of quantum phenomena), or specific three-geometries which do not admit rotational vector fields \cite{sorkin_1980}.

Neutrinos numbers and energies are determined by the temperature at which leptons decoupled from the plasma. At the energies when muons (and hence the neutrinos) decoupled the neutrinos were ultra-relativistic. Both the number, and energy, of the neutrinos are incompatible with models of structure formation \cite{white_83}. By contrast the classical particles would not undergo creation or annihilation reactions and were therefore always decoupled from the plasma (like axions). Classical particles would not be relativistic and would not have a number density constrained by a period of thermal equilibrium with the plasma. The classical particles would however be in gravitational equilibrium, because they would interact gravitationally.

The classical particles would have very similar properties to Black hole relics, BHRs, in the present epoch and are thus an excellent candidate for cold dark matter \cite{macgibbon87}. Unlike BHRs the current density is not constrained by observational limits on Hawking radiation and the density at formation is not restricted by models of black hole formation. It is these two constraints that make primordial black hole relics unlikely candidates for dark matter\cite{chisholm_06} - neither apply to classical particles.

There is little that can be predicted about the mass spectrum. The simple particle-like solutions to the equations of general relativity that we know of are characterised by continuous mass and angular momentum parameters (Kerr and Schwarzschild metrics are well known examples, the Brill and Hartle geon is another family\cite{brill_hartle}). Even negative mass solutions can be written down. However general relativity is a non-liner theory and could also admit discrete solutions. Extensions of general relativity open up more possibilities. If known particles have substantial mass contributions from their interactions - eg the charge contributing to the self energy of the electron, then a classical particle, experiencing only gravitational interactions, would be expected to have masses less than all known particles (except possibly neutrinos).

Classical dark matter, would be distinguished by having purely gravitational interactions. Non gravitational effects arising from the internal structure would take place only at distances of the order $r_m$ which for masses of the order a few eV, would be vanishingly small. The interaction cross sections would be too small for any existing (or conceivable) direct detection experiments. They would be in gravitational equilibrium with other matter as is consistent with galactic rotation curves. The density distribution would not have been perturbed by any non-gravitational effects.

\section{Conclusion}

It may well be found that quantum theory is ubiquitous and that a form of string theory does eventually predict all particle wavefunctions, including those that describe dark matter. But the laws of Nature are not determined by popular vote. With what we know today, classical particles could coexist with quantum particles. If so they would manifest themselves as cold dark matter - consistent with observational evidence and models of structure formation. With the characteristics of primordial black hole relics, but without the number and density constraints imposed by models of black hole formation and evaporation.

\section*{References}


\end{document}